\documentclass{IEEEtran}
\usepackage[normalem]{ulem}
\usepackage{graphicx}
\usepackage{xcolor}
\usepackage{amssymb}

\usepackage{multirow}

\usepackage{multicol}

\usepackage{diagbox}

\usepackage{tabularx}

\usepackage[cmex10]{amsmath}

\usepackage{url}

\usepackage{pifont}

\usepackage{dsfont}
 \usepackage{amsmath}
\usepackage{cite}

\usepackage{float}
\usepackage{subfig}
\usepackage{amsfonts,amssymb}

\usepackage[absolute,showboxes]{textpos}

\TPMargin{3pt}

\newcommand{\copyrightstatement}{
    \begin{textblock}{0.84}(0.08,0.953) 
         \noindent
         \footnotesize
         \copyright 2024 IEEE. Personal use of this material is permitted. Permission from IEEE must be obtained for all other uses, in any current or future media, including reprinting/republishing this material for advertising or promotional purposes, creating new collective works, for resale or redistribution to servers or lists, or reuse of any copyrighted component of this work in other works. DOI: 10.1109/LWC.2024.3360053
    \end{textblock}
}
\begin{document}
\copyrightstatement
\title{Air-to-Ground Cooperative OAM Communications}
%
%
%
\author{Ruirui Chen, Yu Ding, Beibei Zhang, Song Li, and Liping Liang
\thanks{This work is supported by the Fundamental Research Funds for the Central Universities under Grant 2019QNB01. (Corresponding author: Liping Liang.)
\par Ruirui Chen, Yu Ding, Beibei Zhang, and Song Li are with School of Information and Control Engineering, China University of Mining and Technology, Xuzhou, 221116, China (emails: rrchen, yding, lb15060029, lisong@cumt.edu.cn).
Liping Liang is with State Key Laboratory of Integrated Services Networks, Xidian University, Xi'an, 710071, China (e-mail: liangliping@xidian.edu.cn). Beibei Zhang is also with Jiangsu Automation Research Institute, Lianyungang, 222061, China.
}
}

\markboth{IEEE WIRELESS COMMUNICATIONS LETTERS, VOL. 13, NO. 4, APRIL 2024}%
{Shell \MakeLowercase{\textit{et al.}}: Bare Demo of IEEEtran.cls for IEEE Journals}

\maketitle

\begin{abstract}
For users in hotspot region, orbital angular momentum (OAM) can realize multifold increase of spectrum efficiency (SE), and the flying base station (FBS) can rapidly support the real-time communication demand. However, the hollow divergence and alignment requirement impose crucial challenges for users to achieve air-to-ground OAM communications, where there exists the line-of-sight path. Therefore, we propose the air-to-ground cooperative OAM communication (ACOC) scheme, which can realize OAM communications for users with size-limited devices. The waist radius is adjusted to guarantee the maximum intensity at the cooperative users (CUs). We derive the closed-form expression of the optimal FBS position, which satisfies the antenna alignment for two cooperative user groups (CUGs). Furthermore, the selection constraint is given to choose two CUGs composed of four CUs. Simulation results are provided to validate the optimal FBS position and the SE superiority of the proposed ACOC scheme.
\end{abstract}


\IEEEpeerreviewmaketitle

\section{Introduction}

\IEEEPARstart{O}{rbital} angular momentum (OAM) has attracted a lot of interest in wireless communications for high spectrum efficiency (SE) transmission \cite{1}. As a new orthogonal resource, OAM mode produces the helical phase front, which can be utilized for multiplexing communications \cite{2}. The number of OAM mode is theoretically infinite, thus providing vortex wave communications with huge potential for significant SE improvement \cite{3}.

By using the reflector antenna, the authors of \cite{4} successfully separated two vortex wave signals at the same time and same frequency for the first time in 2011. To complete the reception of the vortex wave signal, the general receiving scheme utilized the reflector antenna and centralized antenna array \cite{5}. For  ideal OAM communications, the authors of \cite{6} proposed the mode combination method, which can achieve orthogonal OAM sub-channels at the same time and frequency. To improve information transmission, the authors in \cite{7} combined OAM with index modulation by using different OAM modes. The authors of \cite{8} and \cite{9} enhanced the channel capacity by integrating OAM with non-orthogonal multiple access (NOMA). Note that OAM communications require the line-of-sight (LoS) path, which can not be guaranteed in terrestrial wireless channel that mainly consists of non-LoS (NLoS) paths \cite{10}. The air-to-ground link, which is LoS-dominated, can be rapidly established by the flying base station (FBS) in hotspot region \cite{11}. Thus, we concentrate on OAM communications in the air-to-ground scenario.

Since the wavelength at radio frequency is much larger than that at light frequency, the hollow divergence imposes the critical but challenging problem for air-to-ground OAM communications \cite{12}.
The energy ring of the OAM beam expands as the communication distance increases, thus making the size of the receiving antenna big at the long distance \cite{13}.
In general, the user has size-limited device and only one antenna, thus hardly receiving the vortex wave signal for high SE transmission \cite{14}.
As the easily realized receiving scheme, the cooperative OAM communication scheme was proposed for users in hotspot region to achieve 2 OAM mode multiplexing \cite{15}.
The other challenge for air-to-ground OAM communications is the antenna misalignment, which seriously limits the application of the OAM mode multiplexing.
If the transmitting and receiving antennas are not aligned, it is difficult to decode the vortex wave signals with different OAM modes \cite{16}.
Due to the random distribution of users in hotspot region, the proposed cooperative OAM communication scheme in \cite{17} still has the antenna misalignment scenario, thus requiring beam steering for performance compensation of OAM communications.




To break through the hollow divergence and antenna alignment limitations, we propose the air-to-ground cooperative OAM communication (ACOC) scheme to realize OAM mode multiplexing communications for users with size-limited devices in hotspot region.
The main contributions of this letter can be summarized as follows: 1) We adjust the waist radius to guarantee the maximum intensity at the cooperative users (CUs) according to the distance between the CUs; 2) The optimal FBS position is derived in closed-form solution, which satisfies the alignment requirement of air-to-ground OAM communications; 3) We give the selection constraint to choose two cooperative user groups (CUGs) composed of four CUs.

\section{System Model}
In this letter, there are $U$ users and a FBS in hotspot region due to the concert, stadium game, etc. We denote the user set by $\mathcal{U}=\{1,2\cdots,U\}$. The users all have only one antenna and are randomly distributed in hotspot region.
The FBS has the uniform circular array (UCA), which consists of $N$ antenna elements represented by the set $\mathcal{N}=\{1,2\cdots,N\}$.
Moreover, the equipped UCA has the radius $R$ and central angle $\varphi=2\pi/N$. The channel is assumed as the LoS-dominated and free-space path loss model. The location information of users is available to the FBS.

In the 3D cylindrical coordinate system, the $u$-th $(u\in\mathcal{U})$ user is located at $(r_u, \theta_u, 0)$ and the FBS flies at the height $H$.
The users follow the uniform distribution on the ground ($z=0$) in hotspot region.
The OAM beam is the Laguerre-Gaussian (LG) beam with the radial index $p=0$, the intensity distribution of which can be expressed as follows:
\begin{equation} \label{2}
I_\ell(r,z)=\frac {2}{\pi w^{2}(z)|\ell |!} \left [{\frac {\sqrt {2}r}{w(z)}}\right]^{2|\ell |} \exp \left [{\frac {-2r^{2}}{w^{2}(z)}}\right],
\end{equation}
where $w(z) = {w_0}\sqrt {1 + {{(z/{z_R})}^2}} $ is the radius of the beam at $z$ and the azimuthal index $\ell$ denotes the OAM mode. $w_0$ represents the waist radius at $z=0$. $z_R=\pi w_0^2/\lambda$ is the Rayleigh range, where $\lambda$ denotes the wavelength.

Due to the strict convexity of $I_\ell(r,z)$ with respect to $r$, the optimal radial radius $r$, which corresponds to the maximum $I_\ell(r,z)$, can be derived as
\begin{equation} \label{3}
r_\textup{max}(z)=\sqrt{\frac{|\ell |}{2}}w(z).
\end{equation}
Based on Eq. (\ref{3}) and size limitation of the FBS, we select the smaller waist radius $w_0$, i.e.,
\begin{equation} \label{12ee}
w_0=\sqrt{\frac{r_\textup{max}^2(z)}{|\ell |}- \sqrt{\frac{r_\textup{max}^4(z)}{|\ell |^2}-\frac{z^2\lambda^2}{\pi^2}}}.
\end{equation}
The following condition of the feasible radial radius should be guaranteed for the existence of the waist radius $w_0$.
\begin{equation} \label{11ee}
r_\textup{max}(z) \geq r_\textup{fea}(z) \triangleq \sqrt{\frac{z\lambda |\ell |}{\pi}}
\end{equation}

In general, the user is equipped with size-limited device and only one antenna, thus hardly receiving the vortex wave signal. To achieve $L$ OAM mode multiplexing communications, $L$ antennas should be deployed uniformly alongside the $2\pi/a$ arc and different OAM modes $l_m ~(m=1, 2, 3...M)$ should satisfy the equation $l_m-l_0=ma$, where $l_0$ and $m$ are respectively the constant integer and arbitrary integer \cite{13}. We can see that $L$ CUs can cooperate with each other to achieve $L$ OAM mode multiplexing communications, which requires the CUs to satisfy the uniform distribution alongside the circle. It is hard to realize OAM communications with the large number of OAM modes. Therefore, we choose two CUs as one CUG to achieve air-to-ground cooperative OAM communications.

\section{Air-to-Ground Cooperative OAM Communication Scheme}

We define two CUs as the $u_1$-th and $u_2$-th users, which can be considered as CUG 1. The distance between two CUs is denoted by $d_{{u_1},{u_2}}$. Similarly, CUG 2 consists of the $u_3$-th and $u_4$-th users, the distance between which can be represented by $d_{{u_3},{u_4}}$.
\begin{figure}[thp]
\setlength{\abovecaptionskip}{0.cm}
\setlength{\belowcaptionskip}{-0.cm}
\centering
\vspace{-0.35cm}
\setlength{\abovecaptionskip}{0.cm}
\setlength{\belowcaptionskip}{-0.cm}
\includegraphics[height=2.6in,width=7.8cm]{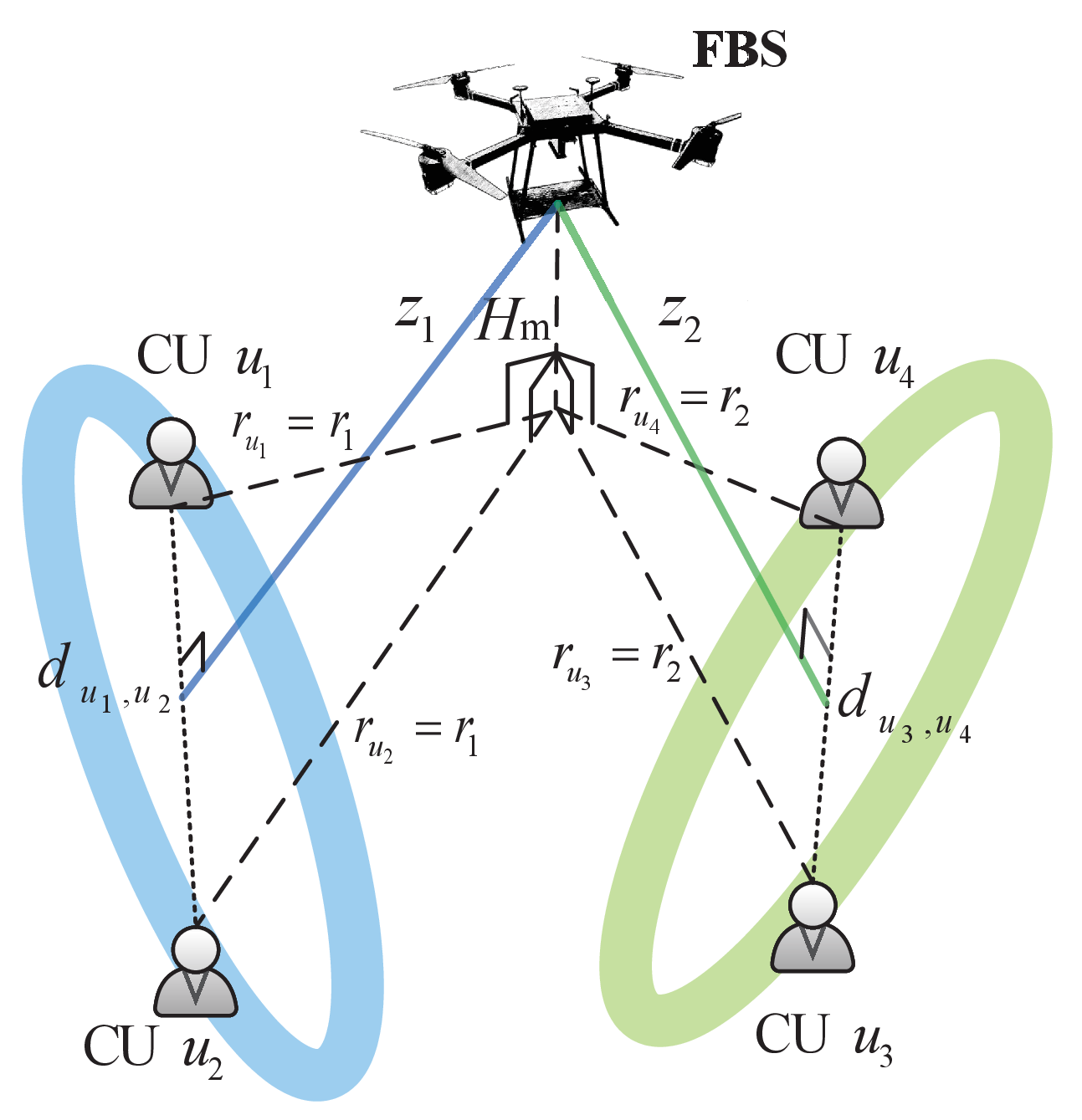}
\caption{\! The selection of two CUGs}
\label{fig:1.1}
\intextsep=1pt plus 3pt minus 1pt
\end{figure}
According to Eq. (\ref{3}), two CUs should be close to the radial radius with maximum intensity. To this end, two CUs should have same distance to the FBS and be at the diameter of the circle, which is shown in Fig. \ref{fig:1.1}. For CUG 1, two CUs should be located at the blue circle with radius $r_\textup{max}(z_1)$, $r_{u_1} =r_{u_2} =r_1 $, and $d_{{u_1},{u_2}}=2r_\textup{max}(z_1)$, where $z_1$ represents the distance from the UCA center of the FBS to the midpoint of the connection between two CUs, which is written as
\begin{equation}\label{5aaab}
{z_1} = \sqrt {r_1^2 -\frac{ d_{u_1, u_2}^2}{4} + {H^2}}.
\end{equation}
The FBS with the height $H$ sends vortex wave signals to CUG 1 through the direction of $z_1$, which is marked by the blue solid line in Fig. \ref{fig:1.1}. Note that ${z_1}$ can be considered as the transmission distance of the vortex wave signal. Based on Eq. (\ref{5aaab}), the transmission distance of the vortex wave signal will increase with the increase of the FBS height $H$. Thus, the FBS should fly at the allowable minimum flying height $H_\textup{m}$.

For the CUs, we define the maximum communication distance as $d_\textup{mc}$.
The waist radius $w_0$ is adjusted by the FBS to guarantee that two CUs are at the diameter of the circle with radius $r_\textup{max}(z_1)$. That is to say, we change the waist radius $w_0$ according to the radial radius $r_\textup{max}(z_1)=d_{u_1, u_2}/2$, and the direction of the OAM beam is from the UCA center of the FBS to the midpoint of the connection between two CUs. According to Eq. (\ref{11ee}), the condition $d_{{u_1},{u_2}}\!\geq\! 2r_\textup{fea}(z_1)$ should be guaranteed to make the waist radius exist.
Similarly, CUG 2 has the propagation direction $z_2$, and should be located at the green circle with radius $r_\textup{max}(z_2)$, $r_{u_3} =r_{u_4} =r_2$, and $d_{{u_3},{u_4}}=2r_\textup{max}(z_2)$.

This letter aims to realize aligned air-to-ground OAM communications for two CUGs by utilizing the mobility of the FBS. To this end, we derive the closed-form expression of the optimal FBS position, which realizes aligned air-to-ground cooperative OAM communications. Furthermore, the selection constraint is given to choose two CUGs composed of four CUs.

\subsection{Optimal FBS Position}

Considering the FBS mobility, the optimal position of the FBS is the midpoint of the connection between two CUs, which can guarantee the minimum transmission distance for one CUG. It is easy for one CUG to obtain the maximum SE by adjusting the position of the FBS and the waist radius of the OAM beam. Therefore, we study the optimal FBS position to realize the air-to-ground cooperative OAM communications for two CUGs composed of four CUs. To guarantee the antenna alignment, the propagation direction for the CUG should be the perpendicular bisector of the connection between two CUs. In theory, the antenna alignment can be guaranteed if the FBS is at the perpendicular bisection plane of the connection between two CUs. To guarantee antenna alignment of two CUGs, the optimal FBS position should be at the height $H_\textup{m}$ and the intersection line of the two perpendicular bisection planes of two CUGs.

\begin{figure}[htbp]
\setlength{\abovecaptionskip}{0.cm}
\setlength{\belowcaptionskip}{-0.cm}
\centering
\vspace{-0.25cm}
\setlength{\abovecaptionskip}{0.cm}
\setlength{\belowcaptionskip}{-0.cm}
\includegraphics[height=2.3in,width=9.2cm]{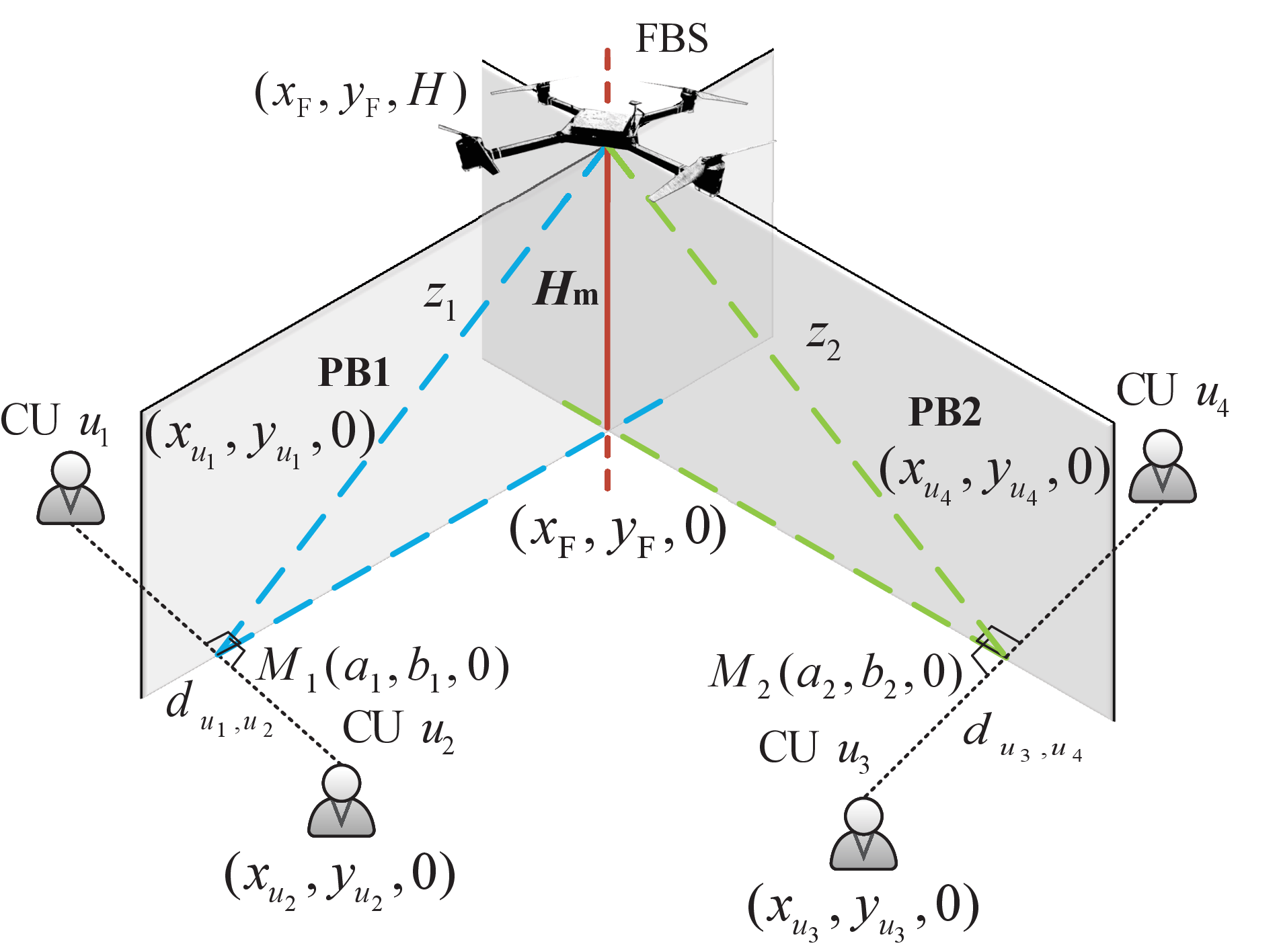}
\caption{ The demonstration of optimal FBS position}
\label{fig:2-1}
\intextsep=1pt plus 3pt minus 1pt
\end{figure}

We give Fig. \ref{fig:2-1} to demonstrate the optimal position of the FBS, which satisfies the alignment requirement of air-to-ground OAM communications. As shown in Fig. \ref{fig:2-1}, we find the perpendicular bisection plane of the connection between $u_1$-th and $u_2$-th CUs, which can be denoted by PB1. Similarly, the perpendicular bisection plane of the connection between $u_3$-th and $u_4$-th CUs can be represented by PB2. Then, we can find the intersection line of the perpendicular bisection planes PB1 and PB2, which is marked by the red line in Fig. \ref{fig:2-1}. Therefore, the optimal position of the FBS is at the height $H_\textup{m}$ and the intersection line of the two perpendicular bisection planes PB1 and PB2, which can be derived in the following.

The location of the $u$-th  $(u \in U)$ user can also be denoted by $({x_u},{y_u},0)$, where ${x_u} = {r_u}\cos {\theta _u}$ and  ${y_u} = {r_u}\sin {\theta _u}$. Similarly, CUs ${u_1}$, ${u_2}$, ${u_3}$, and ${u_4}$ are located at $({x_{{u_1}}},{y_{{u_1}}}, 0)$, $({x_{{u_2}}},{y_{{u_2}}}, 0)$, $({x_{{u_3}}},{y_{{u_3}}}, 0)$, and $({x_{{u_4}}},{y_{{u_4}}}, 0)$, respectively. In Fig. \ref{fig:2-1}, the midpoint of the connection between CUs ${u_1}$ and ${u_2}$ can be represented by ${M_1}$, the location of which is at $(a_1,b_1, 0)$. Similarly, the midpoint of the connection between CUs ${u_3}$ and ${u_4}$ can be denoted by ${M_2}$, the location of which is at $(a_2,b_2, 0)$. The intersection of the perpendicular bisectors can be denoted by $(x_{\textup{F}},y_{\textup{F}}, 0)$, which satisfies the following equations
\begin{equation}
\left\{ \begin{array}{l}
\frac{{y_{\textup{F}} - b_1}}{{x_{\textup{F}} - a_1}} =  - \frac{1}{{{k_1}}}\\
\frac{{y_{\textup{F}} - b_2}}{{x_{\textup{F}} - a_2}} =  - \frac{1}{{{k_2}}},
\end{array} \right.
\end{equation}
where $k_1=\frac{y_{{u_2}}-y_{{u_1}}}{x_{{u_2}}-x_{{u_1}}}$ and $k_2=\frac{y_{{u_4}}-y_{{u_3}}}{x_{{u_4}}-x_{{u_3}}}$.
By solving the equations, we can derive
\begin{equation} \label{eqa7}
\left\{ \begin{array}{l}
x_{\textup{F}} = \frac{{({b_2} - {b_1}){k_1}{k_2} + {a_2}{k_1} - {a_1}{k_2}}}{{{k_1} - {k_2}}}\\
y_{\textup{F}} = \frac{{{b_1}{k_1} - {b_2}{k_2} + {a_1} - {a_2}}}{{{k_1} - {k_2}}}.
\end{array} \right.
\end{equation}
Based on Eq. (\ref{eqa7}), we can obtain the optimal FBS position as the closed-form expression $(x_{\textup{F}},y_{\textup{F}}, H_\textup{m})$, which is at the height $H_\textup{m}$ and the intersection line of the two perpendicular bisection planes PB1 and PB2.

\subsection{CU Selection}

\begin{figure}[thp]
\setlength{\abovecaptionskip}{0.cm}
\setlength{\belowcaptionskip}{-0.cm}
\centering
\vspace{-0.35cm}
\setlength{\abovecaptionskip}{0.cm}
\setlength{\belowcaptionskip}{-0.cm}
\includegraphics[height=2.3in,width=6.9cm]{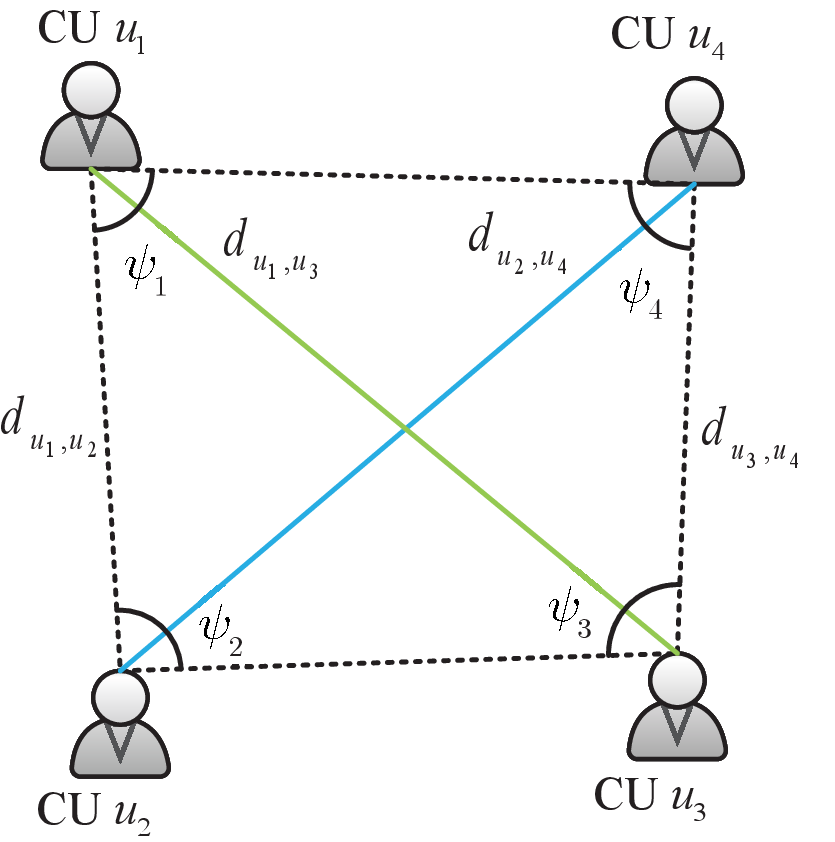}
\caption{\! The planar demonstration of two CUGs}
\label{fig:1.2}
\intextsep=1pt plus 3pt minus 1pt
\end{figure}

There exists the optimal position of the FBS if four CUs $u_1$, $u_2$, $u_3$, and $u_4$ are not at one straight line. Note that it is a very small probability that four CUs are at one straight line. Therefore, four CUs can almost always be selected to form two CUGs in general. In Fig. \ref{fig:1.2}, we give the planar demonstration of two CUGs, which is a quadrilateral with vertices $u_1$, $u_2$, $u_3$, and $u_4$. The diagonals of the quadrilateral can be represented by $d_{u_1, u_3}$ and $d_{u_2, u_4}$, which should satisfy the following condition
\begin{equation}\label{eq8}
\textup{max}\{d_{u_1, u_3}, d_{u_2, u_4}\} \leq 2r_{\rm{sr}},
\end{equation}
where $r_{\rm{sr}}$ is the FBS service radius. Eq. (\ref{eq8}) can guarantee that four CUs can be served by the FBS. The distances of two CUGs should satisfy the following conditions
\begin{equation}\label{eq9}
2r_{\rm{fea}}(z_1) \leq d_{u_1, u_2} \leq d_{\rm{mc}}
\end{equation}
and
\begin{equation}\label{eq10}
2r_{\rm{fea}}(z_2) \leq d_{u_3, u_4} \leq d_{\rm{mc}},
\end{equation}
which can guarantee the efficient information transmission for two CUGs. Based on the above analysis, two CUGs composed of four CUs, which satisfy Eqs. (\ref{eq8}), (\ref{eq9}), and (\ref{eq10}), can be chosen to complete aligned air-to-groud cooperative OAM communications. Therefore, Eqs. (\ref{eq8}), (\ref{eq9}), and (\ref{eq10}) can be considered as the selection constraint to choose two CUGs composed of four CUs.

The inner angle set of the quadrilateral composed by two CUGs can be expressed as
\begin{equation}
\Psi=\{\psi_1, \psi_2, \psi_3, \psi_4\}.
\end{equation}\
For the inner angle set $\Psi$, the angle square difference can be given in the following.
\begin{equation} \label{13eaa}
\overline{\Psi}=\sum^{4}_{i=1}(\psi_i-\frac{\pi}{2})^2
\end{equation}
To further improve the SE, we should choose two CUGs composed of four CUs, which have the minimum $\overline{\Psi}$ and satisfy Eqs. (\ref{eq8}), (\ref{eq9}), and (\ref{eq10}).
\subsection{Scheme Description}
For CU selection, the first CU $u_1$ can be the boundary user that has the maximum $r_u$. The CUs $u_2$, $u_3$, and $u_4$ can be chosen as the users closest, second closest and third closest to CU $u_1$. We define the selected CU set as $\mathcal{U}_{\rm{c}}$.
Then, we propose the ACOC scheme, which can be summarized as follows.

\par \noindent\rule[0.25\baselineskip]{9cm}{0.5pt}
\uline{\textbf{Air-to-Ground Cooperative OAM Communication Scheme}}
\par \noindent\emph{\textbf{Initialization}}:
\par \noindent~1) Given user set $\mathcal{U}$, user location information $(r_u, \theta_u, 0)$
\par \noindent~~~~~and arbitrarily small value $\epsilon$;
\par \noindent~2) Set $\overline{\Psi}=+\infty$, $\mathcal{U}_{\rm{c}}=\emptyset$ and choose boundary user as CU
\par \noindent~~~~~$u_1$ in $\mathcal{U}$;
\par \noindent\emph{\textbf{Iteration}}:
\par \noindent~1) \textbf{while} $(\overline{\Psi}>\epsilon)$\&($\mid\mathcal{U}\mid>3$) \textbf{do}
\par \noindent~2) ~~Find CUs $u_2$, $u_3$, $u_4$ and obtain $d_{u_1, u_2}$, $d_{u_3, u_4}$;
\par \noindent~3) ~~~~\textbf{if} ($2r_{\rm{fea}} \leq d_{u_1, u_2} \leq d_{\rm{mc}})$\&$(2r_{\rm{fea}} \leq d_{u_3, u_4} \leq d_{\rm{mc}})$
\par \noindent~4) ~~~~~~~Calculate $d_{u_1, u_3}$ and $d_{u_2, u_4}$;
\par \noindent~5) ~~~~~~~\textbf{if} ($\textup{max}\{d_{u_1, u_3}, d_{u_2, u_4}\} \leq 2r_{\rm{sr}}$)
\par \noindent~6) ~~~~~~~~~~Obtain $\overline{\Psi}_{\rm{in}}$ based on Eq. (\ref{13eaa});
\par \noindent~7) ~~~~~~~~~~\textbf{if} $(\overline{\Psi}>\overline{\Psi}_{\rm{in}}$)
\par \noindent~8) ~~~~~~~~~~~~~Set $\overline{\Psi}=\overline{\Psi}_{\rm{in}}$ and $\mathcal{U}_{\rm{c}}=\{ u_1, u_2, u_3, u_4 \}$;
\par \noindent~9) ~~~~~~~~~~\textbf{end if}
\par \noindent10) ~~~~~~~\textbf{end if}
\par \noindent11) ~~~~\textbf{end if}
\par \noindent12) ~~Set $\mathcal{U}=\mathcal{U} \setminus u_1$ and $u_1=u_2$;
\par \noindent13) \textbf{end while}
\par \noindent14) According to locations of selected four CUs, obtain
\par \noindent~~~~~optimal FBS position $(x_{\textup{F}},y_{\textup{F}}, H_\textup{m})$ based on Eq. (\ref{eqa7}).
\par \noindent\rule[0.25\baselineskip]{9cm}{1pt}

In initialization steps, we give the arbitrarily small value $\epsilon$, and choose boundary user as CU $u_1$ in $\mathcal{U}$. For each iteration, we choose CUs $u_2$, $u_3$, and $u_4$, which are the users closest, second closest and third closest to CU $u_1$ (step 2 of iteration). If the distances $d_{u_1, u_2}$ and $d_{u_3, u_4}$ both belong to $[2r_{\rm{fea}}, d_{\rm{mc}}]$, we calculate $d_{u_1, u_3}$ and $d_{u_2, u_4}$ (steps 3-4 of iteration). $\overline{\Psi}_{\rm{in}}$ is calculated based on Eq. (\ref{13eaa}) when the four CUs all can be served by the FBS (steps 5-6 of iteration). If $\overline{\Psi}_{\rm{in}}$ is smaller than the current angle square difference, we update the four CUs and set $\overline{\Psi}_{\rm{in}}$ as the current angle square difference (steps 7-8 of iteration). Then, we remove the CU $u_1$ from $\mathcal{U}$, and set $u_2$ as new $u_1$ in the next iteration (step 12 of iteration). When the iteration terminates, we can obtain four CUs in $\mathcal{U}_{\rm{c}}$. The optimal FBS position is at the height $H_\textup{m}$ and the intersection line of the two perpendicular bisection planes of the two CUGs (step 14 of iteration).

\section{Simulation Results}

In this section, simulation results are provided to evaluate the performance of our proposed ACOC scheme. The users follow the uniform distribution in hotspot region that is the square area. For performance comparison, we give the existing cooperative OAM wireless (COW) communication scheme, where the CUs are served by the ground BS. Note that our proposed ACOC scheme can obtain the optimal FBS position to realize the antenna alignment. The simulation parameter is shown in Table I.

\begin{table}[htbp]
 \centering
\caption{Simulation Parameter}
 \label{table1}
\setlength{\tabcolsep}{2.2mm}{
 \begin{tabular}{|c|c|}
  \hline
       &  \\[-5pt]
  \textbf{Parameter}      &   \textbf{Numerical Value}    \\
  \hline
       &  \\[-5pt]
  Carrier Frequency       &  1 GHz            \\
        \hline
       &  \\[-5pt]
  FBS Transmit Power &    30 dBm           \\
        \hline
        &  \\[-5pt]
  Noise Power        &   -90 dBm             \\
        \hline
        &  \\[-5pt]
  OAM Mode             & 1       \\
        \hline
        &  \\[-5pt]
  Height of FBS         &   50-250 m             \\
        \hline
               &  \\[-5pt]
  Height of Ground BS         &   50-250 m             \\
        \hline
 \end{tabular}}

\end{table}

\begin{figure}[htbp]
\setlength{\abovecaptionskip}{0.cm}
\setlength{\belowcaptionskip}{-0.cm}
\centering
\vspace{-0.25cm}
\setlength{\abovecaptionskip}{0.cm}
\setlength{\belowcaptionskip}{-0.cm}
\includegraphics[height=2.4in,width=9cm]{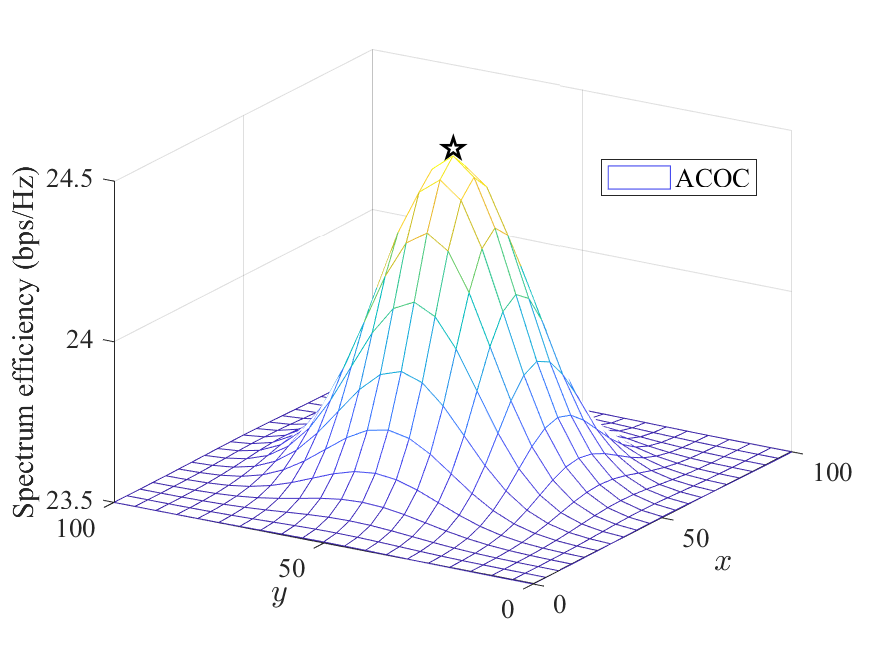}
\caption{ The variation of the SE with the FBS position}
\label{fig:5}
\intextsep=1pt plus 3pt minus 1pt
\end{figure}

Figure \ref{fig:5} presents the variation of the SE with the FBS position. The FBS flies at the allowable minimum flying height $H_\textup{m}$. The optimal FBS position $(x_{\textup{F}},y_{\textup{F}}, H_\textup{m})$ is marked by \ding {73}. We can see that the SE achieves the largest value at the optimal FBS position $(x_{\textup{F}},y_{\textup{F}}, H_\textup{m})$, which can validate the closed-form expression of the optimal FBS position in Eq. (\ref{eqa7}).

\begin{figure}[h]
	
	\begin{minipage}{0.49\linewidth}
		\vspace{3pt}
		\centerline{\includegraphics[height=1.6in,width=4.4cm]{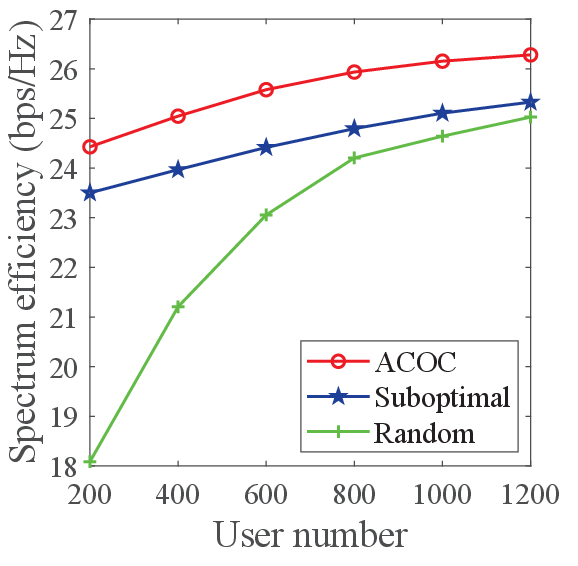}}
		\centerline{(a)}
	\end{minipage}
	\begin{minipage}{0.49\linewidth}
		\vspace{3pt}
		\centerline{\includegraphics[height=1.6in,width=4.5cm]{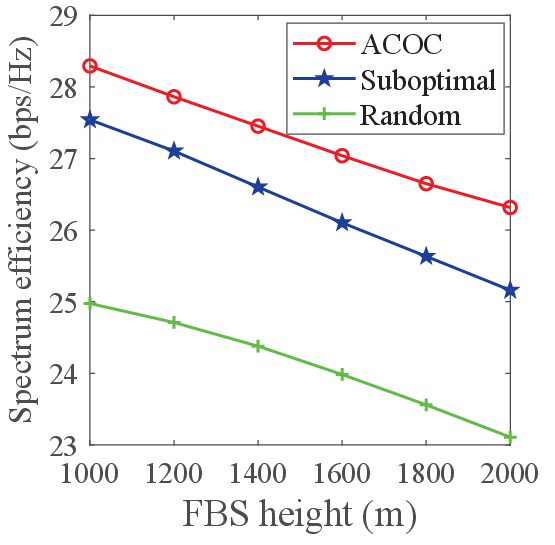}}
	
		\centerline{(b)}
	\end{minipage}
	\caption{The SE comparison}
	\label{fig:6}
\end{figure}

In Fig. \ref{fig:6}, we compare the proposed ACOC shceme with the suboptimal scheme and random scheme. The FBS position of the suboptimal scheme only guarantees the antenna alignment of one CUG. The random scheme sets the random FBS position. The FBS height and user number are 50 m and 4000 in Fig. \ref{fig:6}(a) and Fig. \ref{fig:6}(b), respectively. For the three schemes, the SE monotonically increases/decreases as the user number/FBS height increases. The proposed ACOC scheme obtains higher SE than the suboptimal scheme and random scheme. This is due to the fact that the optimal FBS position, which satisfies alignment requirement, can achieve aligned air-to-ground OAM communications for two CUGs.
\begin{figure}[htbp]
\setlength{\abovecaptionskip}{0.cm}
\setlength{\belowcaptionskip}{-0.cm}
\centering
\vspace{-0.25cm}
\setlength{\abovecaptionskip}{0.cm}
\setlength{\belowcaptionskip}{-0.cm}
\includegraphics[height=2.8in,width=9.0cm]{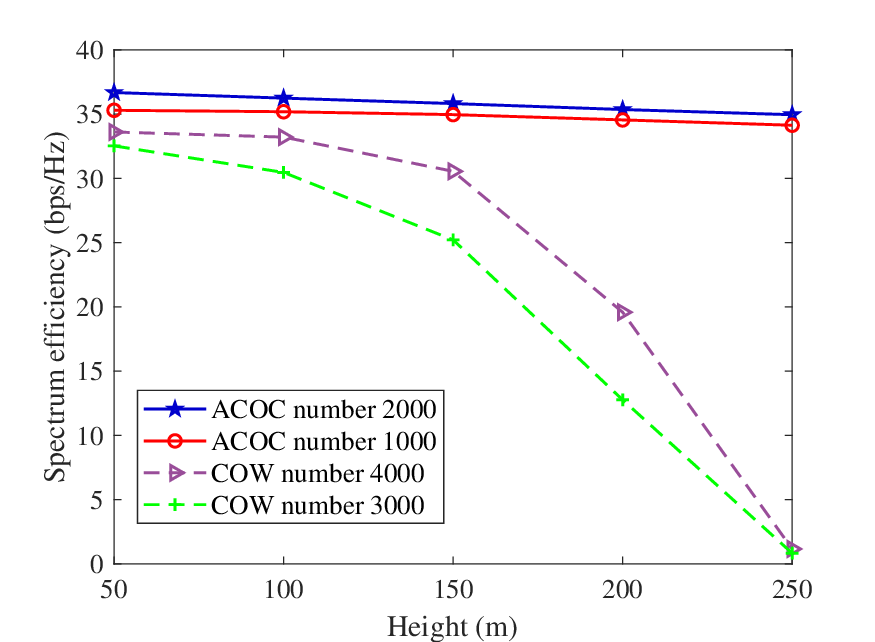}
\caption{ The SE versus height}
\label{fig:2}
\intextsep=1pt plus 3pt minus 1pt
\end{figure}

Figure \ref{fig:2} presents the SE performance under various heights. From Fig. \ref{fig:2}, we can see that the larger the user number is, the larger the SE is for the two schemes. Furthermore, the SE is a monotonic decreasing function of the FBS height. This is due to the fact that as the FBS height increases, the transmission distance of the vortex wave signal will increase, thus making the minimum communication distance between two CUs larger. With less user number, the proposed ACOC scheme still can obtain higher SE than that of the COW communication scheme.

\begin{figure}[htbp]
\setlength{\abovecaptionskip}{0.cm}
\setlength{\belowcaptionskip}{-0.cm}
\centering
\vspace{-0.25cm}
\setlength{\abovecaptionskip}{0.cm}
\setlength{\belowcaptionskip}{-0.cm}
\includegraphics[height=2.8in,width=9.0cm]{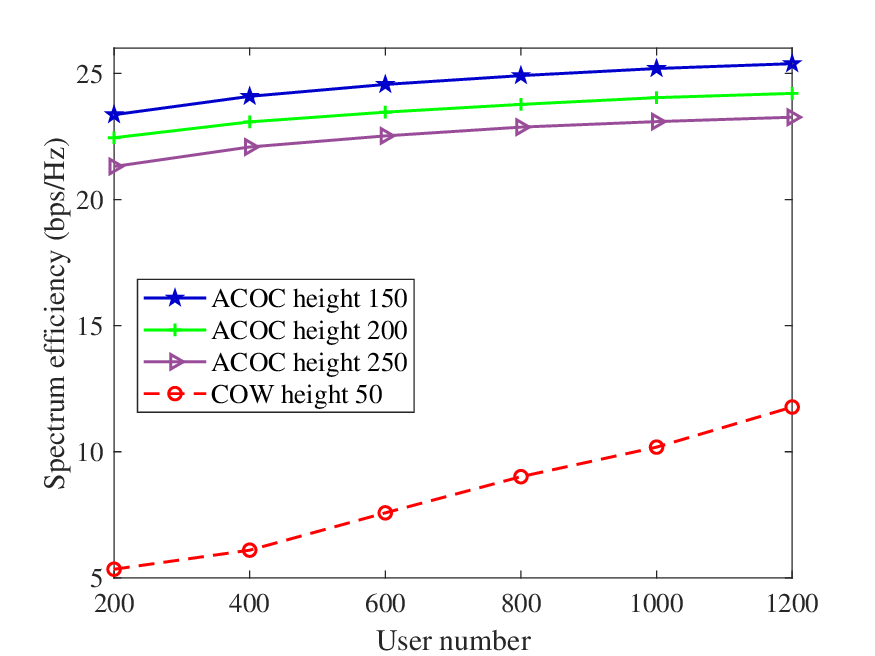}
\caption{Proposed ACOC scheme under various user numbers}
\label{fig:3}
\intextsep=1pt plus 3pt minus 1pt
\end{figure}

In Fig. \ref{fig:3}, we demonstrate the SE performance versus user number. For the two schemes, it can be observed that the SE monotonically increases as the user number increases. This is because as the user number increases, the two schemes both obtain the large probability to realize long distance information transmission. Compared with the COW communication scheme, the proposed ACOC scheme achieves better SE performance even with higher position.

\section{Conclusions}

To overcome the hollow divergence and antenna alignment limitations, this letter studied the ACOC scheme to realize air-to-ground OAM communications for users with size-limited devices in hotspot region. The waist radius was adjusted to make the CUs at the maximum intensity, which can guarantee the reception of vortex wave signal. The optimal FBS position was derived in closed-form solution to achieve aligned air-to-ground cooperative OAM communications. Furthermore, we give the selection constraint to choose two CUGs composed of four CUs. Simulation results showed that the proposed ACOC scheme can achieve high SE transmission at the long distance.


\begin{thebibliography}{99}
\bibitem{1}
W. Cheng, W. Zhang, H. Jing, et al., ``Orbital angular momentum for wireless communications," \emph{IEEE Wireless Communications}, vol. 26, no. 1, pp. 100-107, Feb. 2019.
\bibitem{2}
X. Ge, R. Zi, X. Xiong, et al., ``Millimeter wave communications with OAM-SM scheme for future mobile networks," \emph{IEEE Journal on Selected Areas in Communications}, vol. 35, no. 9, pp. 2163-2177, Sept. 2017.
\bibitem{3}
W. Yu, B. Zhou, Z. Bu, et al., ``UCA based OAM beam steering with high mode isolation," \emph{IEEE Wireless Communications Letters}, vol. 11, no. 5, pp. 977-981, May 2022.
\bibitem{4}
F. Tamburini, E. Mari, A. Sponselli, et al., ``Encoding many channels on the same frequency through radio vorticity: first experimental test," \emph{New Journal Of Physics}, vol. 14, no. 3, Art. 2012.
\bibitem{5}
R. Chen, H. Zhou, M. Moretti, et al., ``Orbital angular momentum waves: generation, detection, and emerging applications," \emph{IEEE Communications Surveys and Tutorials}, vol. 22, no. 2, pp. 840-868, 2020.
\bibitem{6}
K. A. Opare, Y. Kuang and J. J. Kponyo, ``Mode combination in an ideal wireless OAM-MIMO multiplexing system," \emph{IEEE Wireless Communications Letters}, vol. 4, no. 4, pp. 449-452, Aug. 2015.
\bibitem{7}
E. Basar, ``Orbital angular momentum with index modulation," \emph{IEEE Transactions on Wireless Communications}, vol. 17, no. 3, pp. 2029-2037, Mar. 2018.
\bibitem{8}
A. A. Amin and S. Y. Shin, ``Capacity analysis of cooperative NOMA-OAM-MIMO based full-duplex relaying for 6G," \emph{IEEE Wireless Communications Letters}, vol. 10, no. 7, pp. 1395-1399, Jul. 2021.
\bibitem{9}
A. A. Amin and S. Y. Shin, ``Channel capacity analysis of non-orthogonal multiple access with OAM-MIMO system," \emph{IEEE Wireless Communications Letters}, vol. 9, no. 9, pp. 1481-1485, Sept. 2020.
\bibitem{10}
X. Pang, M. Liu, N. Zhao, et al., ``Secrecy analysis of UAV-based mmWave relaying networks," \emph{IEEE Transactions on Wireless Communications}, vol. 20, no. 8, pp. 4990-5002, Aug. 2021.
\bibitem{11}
R. Chen, Y. Sun, L. Liang, et al., ``Joint power allocation and placement scheme for UAV-assisted IoT with QoS guarantee," \emph{IEEE Transactions on Vehicular Technology}, vol. 71, no. 1, pp. 1066-1071, Jan. 2022.
\bibitem{12}
R. Chen, J. Zhou, W. Long, et al., ``Hybrid circular array and luneberg lens for long-distance OAM wireless communications," \emph{IEEE Transactions on Communications}, vol. 71, no. 1, pp. 485-497, Jan. 2023.
\bibitem{13}
K. Liu, H. Liu, Y. Qin, et al., ``Generation of OAM beams using phased array in the microwave band," \emph{IEEE Transactions on Antennas and Propagation}, vol. 64, no. 9, pp. 3850-3857, Sept. 2016.
\bibitem{14}
R. Chen, W. Cheng, Y. Ding, et al., ``QoS-guaranteed multi-UAV coverage scheme for IoT communications with interference management," \emph{IEEE Internet of Things Journal}, 2023. Available:https://ieeexplore.ieee.org/document/10201847.
\bibitem{15}
R. Chen, W. Cheng, J. Lin, et al., ``Cooperative orbital angular momentum wireless communications," \emph{IEEE Transactions on Vehicular Technology}, 2023. Available:https://ieeexplore.ieee.org/document/10232879.
\bibitem{16}
X. Xiong, H. Lou and X. Ge, ``Modeling and optimization of OAM-MIMO communication systems with unaligned antennas," \emph{IEEE Transactions on Communications}, vol. 70, no. 6, pp. 3682-3694, Jun. 2022.
\bibitem{17}
W. Zhang, S. Zheng, Y. Chen, et al., ``Orbital angular momentum-based communications with partial arc sampling receiving," \emph{IEEE Communications Letters}, vol. 20, no. 7, pp. 1381-1384, Jul. 2016.
\end{thebibliography}
\end{document}